\documentstyle[12pt,epsf]{article}
\newcommand{\ba}{\begin{array}}
\newcommand{\ea}{\end{array}}
\newcommand{\bi}{\begin{itemize}}
\newcommand{\ei}{\end{itemize}}
\newcommand{\no}{\nonumber}

\newcommand{\Z}{{\bf Z}}

\newcommand{\R}{{\bf R}}
\newcommand{\C}{{\bf C}}

\newcommand{\cH}{{\cal H}}

\newcommand{\cL}{{\cal L}}

\newcommand{\cN}{{\cal N}}
\newcommand{\cO}{{\cal O}}

\newcommand{\na}{\nabla}

\newcommand{\del}{\partial}

\newcommand{\bra}{\langle}
\newcommand{\ket}{\rangle}

\newcommand{\diag}{\mbox{diag}\,}

\newcommand{\Tr}{{\rm Tr}}

\newcommand{\pr}{\prime}

\newcommand{\rar}{\rightarrow}

\newcommand{\ti}{\times}

\newcommand{\ot}{\otimes}

\newcommand{\unl}{\underline}

\newcommand{\fr}{\frac}
\newcommand{\half}{\frac{1}{2}}
\newcommand{\qua}{\frac{1}{4}}
\newcommand{\scr}{\scriptsize}
\newcommand{\dis}{\displaystyle}
\newcommand{\mn}{{\mu\nu}}
\newcommand{\zb}{\bar{z}}
\newcommand{\ah}{\hat{a}}
\newcommand{\fh}{\hat{f}}
\newcommand{\nh}{\hat{n}}
\newcommand{\uh}{\hat{u}}
\newcommand{\vh}{\hat{v}}
\newcommand{\xh}{\hat{x}}
\newcommand{\zh}{\hat{z}}
\newcommand{\Ah}{\hat{A}}
\newcommand{\Bh}{\hat{B}}
\newcommand{\Dh}{\hat{D}}
\newcommand{\Fh}{\hat{F}}
\newcommand{\Ph}{\hat{P}}
\newcommand{\Uh}{\hat{U}}
\newcommand{\Vh}{\hat{V}}
\newcommand{\Phih}{\hat{\Phi}}
\newcommand{\zbh}{\hat{\bar{z}}} 
\newcommand{\nah}{\hat{\nabla}}
\newcommand{\delh}{\hat{\partial}}

\makeatletter
\@addtoreset{equation}{section}

\makeatother
\begin{document}

\begin{titlepage}
\null
\begin{flushright}
UT-933
\\
YITP-01-63
\\
hep-th/0109070
\\
September, 2001
\end{flushright}

\vskip 1.3cm
\begin{center}
 

   {\LARGE ADHM/Nahm Construction of Localized Solitons}
\vskip 5mm
   {\LARGE Noncommutative Gauge Theories}

\vskip 1.5cm
\normalsize

  {\large Masashi H\small{AMANAKA}\footnote{e-mail:
 hamanaka@hep-th.phys.s.u-tokyo.ac.jp}}

\vskip 1.2cm

  {\large \it Department of Physics, University of Tokyo,\\
               Tokyo 113-0033, Japan}

\vskip 0.6cm

 {\large \it Yukawa Institute for Theoretical Physics \\
             Kyoto University, Kyoto 606-8502, Japan\footnote
   {The author stays at YITP as an atom-type visitor from August 27 to 
             September 20, 2001 and from January 28 to February 2, 2002.}} 

\vskip 1.2cm

{\bf Abstract}
\end{center}

We study the relationship between ADHM/Nahm construction and
``solution generating technique'' of BPS solitons
in noncommutative gauge theories.
ADHM/Nahm construction and
``solution generating technique''
are the most strong ways to construct 
exact BPS solitons.
Localized solitons are the solitons which are generated
by the ``solution generating technique.''
The shift operators which play crucial roles
in ``solution generating technique''
naturally appear in ADHM/Nahm construction
and we can construct various  exact localized solitons 
including new solitons:
localized periodic instantons (=localized calorons) and
localized doubly-periodic instantons.
Nahm construction also gives rise to BPS fluxons straightforwardly 
from the appropriate input Nahm data which is expected from
the D-brane picture of BPS fluxons.
We also show that 
the Fourier-transformed soliton of the localized caloron
in the zero-period limit
exactly coincides with the BPS fluxon.

\end{titlepage}

\clearpage

\baselineskip 6.5mm

\section{Introduction}

Noncommutative gauge theories are fascinating generalizations of 
ordinary gauge theories and often appear mysteriously in string theories. 
Recently, it is shown that
gauge theories on D-branes with background constant $B$-field
are equivalent to noncommutative gauge theories
in some limit \cite{CoDoSc}, \cite{DoHu}, \cite{SeWi} 
and it becomes possible to study some aspects of 
D-brane dynamics such as tachyon condensations\footnote{For a 
review see \cite{Harvey}.}
in terms of noncommutative gauge theories
which is comparatively easier to deal with.
Especially noncommutative BPS solitons are worth studying
because they describe the static configurations of D-branes
and are important in studying non-perturbative
aspects of the gauge theories on it. 

Noncommutative spaces are characterized by the noncommutativity of
the spatial coordinates:
\begin{eqnarray}
\label{nc_coord}
[x^i,x^j]=i\theta^{ij}.
\end{eqnarray}
This relation looks like the canonical commutation 
relation in quantum mechanics
and leads to ``space-space uncertainty relation.''
Hence
the singularity which exists on commutative spaces could resolve
on noncommutative spaces.
This is one of the distinguished features of noncommutative theories
and gives rise to various new physical objects,
for example, smooth $U(1)$ 
instantons \cite{NeSc}, \cite{Fu}\footnote{On commutative side, e.g.
\cite{SeWi}, \cite{MaMiMoSt}, \cite{Tera}, \cite{Moriyama}}, 
``visible Dirac-like strings'' \cite{GrNe} and 
the fluxons \cite{Po}, \cite{GrNe2}. $U(1)$ instantons 
exist due to the resolution of
small instanton singularities of the complete instanton moduli space
\cite{Nakajima}.
However $U(1)$ instantons still exist 
even when the singularities
of the complete instanton moduli space don't resolve,
that is, when the
self-duality of the gauge field 
is the same as that of the noncommutative parameter $\theta^{ij}$
\cite{AgGoMiSt}, \cite{Nek}, \cite{Fu4}.

There are two powerful ways to construct exact noncommutative BPS
solitons, that is, ADHM/Nahm construction 
and ``solution generating technique.''
ADHM or Nahm construction is a wonderful application 
of the one-to-one correspondence
between the instanton or monopole moduli 
space and the space of ADHM or Nahm data
and gives rise to arbitrary instantons \cite{AtHiDrMa}
or monopoles \cite{Nahm} respectively.
ADHM/Nahm construction 
has a remarkable D-brane description
\cite{Witten}, \cite{Douglas}, \cite{Diaconescu}.
D-branes give intuitive explanations for
various results of known field theories and
explain the reason why the instanton or monopole moduli 
spaces and the space of ADHM or Nahm data correspond one-to-one.
However there still exist unknown parts of the D-brane
descriptions and  
it is expected that further study of the D-brane description of 
ADHM/Nahm construction would reveal new aspects of D-brane dynamics, 
such as Myers effect \cite{Myers}
which in fact corresponds to some boundary conditions in Nahm construction.
On the other hand, 
``solution generating technique''
is a transformation which leaves an equation as it is
and gives rise to various new solutions from known solutions of it.
The new solutions have a clear matrix theoretical interpretation 
\cite{BaFiShSu}, \cite{IsKaKiTs}, \cite{AoIsIsKaKiTa}
and concerns with the important fact 
that a D-brane can be constructed by lower dimensional 
D-branes.
Hence the study of the relation between the two constructions
is very important to deepen our understanding of D-branes.

The $U(1)$ instantons with the same self-duality as the noncommutative 
parameter $\theta^{ij}$ and BPS fluxons
can be constructed by applying ``solution generating
technique'' \cite{HaKrLa} to the corresponding BPS equations
\cite{AgGoMiSt} and \cite{GrNe2}, \cite{HaTe}, \cite{Hashi}
respectively\footnote{``Solution generating
technique'' can be also applied to the self-dual BPS equation
of $(2+1)$-dimensional Abelian-Higgs model only when the Higgs vacuum
expectation value $v$ satisfies $v^2=1/\theta$ \cite{Bak},
\cite{HaTe}, \cite{Hashi}, \cite{BaLePa}.}.
The solitons which are generated from the vacuum 
by ``solution generating technique''
are called {\it localized solitons}
in the matrix theoretical contexts.
In general, the new solitons generated from known solitons 
by ``solution generating technique'' are
the composite of known solitons and localized solitons.
Hence localized solitons are essential in 
``solution generating technique''
and, in fact, special to noncommutative gauge theories.
Localized instantons have been constructed 
not only by ``solution generating technique'' \cite{AgGoMiSt}
but also by ADHM construction
\cite{Nek}, \cite{Fu4}.
BPS fluxons are the special class of 
BPS solitons in $(3+1)$-dimensional
noncommutative gauge theory
and must be found by Nahm construction.
However they have not been found yet.
Moreover in order to get BPS fluxons by ``solution generating technique,''
we have to modify the technique \cite{HaTe} or use some trick \cite{Hashi}.

There is another BPS soliton to which ADHM/Nahm construction
can be applied: the caloron.
Calorons are periodic instantons in one direction, 
that is, instantons on $\R^3\times S^1$.
They were first constructed explicitly in \cite{HaSh}
as infinite number of 't Hooft instantons 
periodic in one direction
and used for the discussion on non-perturbative aspects of
finite-temperature field theories \cite{HaSh}, \cite{GrPiYa}.
Calorons can intermediate between instantons and monopoles 
and coincide with them in the limits
of $\beta\rar \infty$ and $\beta\rar 0$ respectively 
where $\beta$ is the perimeter of $S^1$ \cite{Ro}.
Hence calorons also can be 
reinterpreted clearly from D-brane picture \cite{LeYi}
and constructed by Nahm construction 
\cite{Nahm2}, \cite{KrvB}, \cite{LeLu}. 

\begin{center}
\begin{figure}
\epsfxsize=110mm
\hspace{3cm} 
\epsffile{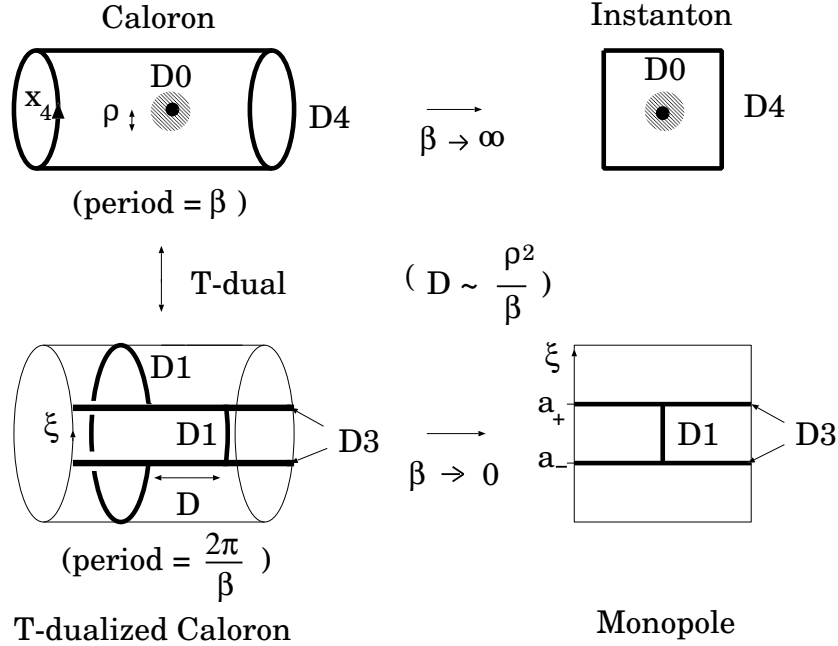}
\caption{The D-brane description of $U(2)$ 1 caloron.}
\label{caloron}
\end{figure}
\end{center}

The D-brane pictures of them are the following (see figure \ref{caloron}).
Instantons and monopoles are represented as D0-branes on D4-branes
and D-strings ending to D3-branes respectively.
Hence calorons are represented as D0-branes on D4-branes
lying on $\R^3\ti S^1$.

\begin{center}
\begin{figure}
\epsfxsize=120mm
\hspace{3cm} 
\epsffile{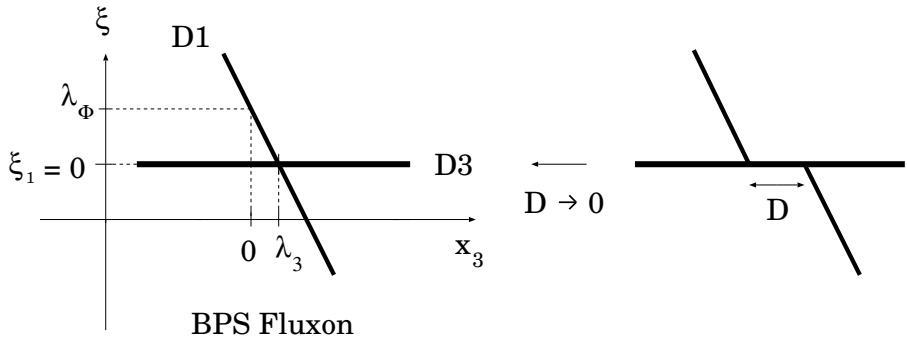}
\caption{The BPS fluxon}
\label{fluxonfig}
\end{figure}
\end{center}

In the T-dualized picture, $U(N)$ 1 caloron can be interpreted
as $N-1$ fundamental monopoles and the $N$-th monopole which appears 
from the Kaluza-Klein sector \cite{LeYi}.
The value of the fourth component of the gauge field
at spatial infinity
on D4-brane determines the positions of the D3-branes
which denote the Higgs expectation values of the monopole.
The positions of the D3-branes are called the jumping points
because at these points, the D1-brane is generally separated.
In $N=2$ case, the separation interval (see figure \ref{caloron}) $D$
satisfies $D\sim \rho^2/\beta$ \cite{LeYi}, \cite{LeLu},
and if the size $\rho$ of periodic instanton is fixed and
the period $\beta$ goes to zero,
then one monopole decouples and 
the situation exactly coincides with that of PS-monopole \cite{PrSo}.
BPS fluxons are represented as infinite D-strings piercing D3-branes
in the background constant $B$-field 
and considered to be the T-dualized noncommutative calorons 
in the limit with the period $\beta\rar0$
and the interval $D\rar 0$, which suggests $\rho =0$. 
(cf. figure \ref{fluxonfig})

In the present paper, 
we give the various exact BPS solitons by ADHM/Nahm construction:
localized instantons,
localized calorons, localized doubly-periodic instantons 
and BPS fluxons which are essential in
``solution generating technique.''
The shift operators which 
play crucial roles in ``solution generating
technique'' naturally appear in ADHM construction
and other important points are all derived straightforwardly
in ADHM/Nahm construction.
In this way, we discuss the relationship
between the two methods.
The solutions of the localized calorons 
and the localized doubly-periodic instantons
are new results.
We also discuss a Fourier transformation 
of the localized calorons
and show that 
the Fourier-transformed configurations of the
localized calorons in the $\beta\rar 0$ limit indeed 
coincides with BPS fluxons, which could be considered that 
BPS fluxons corresponding to D1-branes 
are the solitons of T-dualized solitons
of localized calorons corresponding to D0-brane 
with the period $\beta\rar 0$
up to space rotation \cite{HaHi}, \cite{Moriyama2}, \cite{HaHiMo}.

This paper is organized as follows.
In section 2, we briefly review 
``solution generating technique'' and localized solitons.
In section 3, we present ADHM construction of instantons and 
apply them to localized solitons.
In section 4, we take the Fourier-transformation of the localized calorons
and show that in the $\beta\rar 0$ limit, the transformed solitons
exactly coincide with BPS fluxons.
Finally section 5 is devoted to conclusion and discussion.

\section{A Review of ``Solution Generating Technique'' and Localized Solitons}

In this section, we make a brief review of
``solution generating technique'' and some application of it 
which generates localized instantons and BPS fluxons.

Noncommutative gauge theories have two equivalent descriptions,
that is, star-product formalism and operator formalism.
There is a commutative description equivalent to 
the noncommutative gauge theories and the commutative and the
noncommutative description are connected
by Seiberg-Witten map \cite{SeWi}. 
In the present paper, we mainly use the operator formalism
and when we make a physical interpretation, 
we shift to the commutative description by Seiberg-Witten map.

Let's present noncommutative gauge theories
in the operator formalism and establish notations.
In this formalism, 
we start with the noncommutativity of the spatial coordinates
(\ref{nc_coord}) and define noncommutative gauge theories considering 
the coordinates as operators. 
From now on, we denote the hat on the operators 
in order to emphasize that they are operators.
Here, for simplicity, we treat a noncommutative plane
with the coordinates $\xh^1,\xh^2$ which satisfy $[\xh^1,\xh^2]
=i\theta,~\theta>0$.

Defining new variables $\ah,\ah^\dagger$ as
\begin{eqnarray}
\ah:=\fr{1}{\sqrt{2\theta}}\zh,~\ah^\dagger:=\fr{1}{\sqrt{2\theta}}\zbh,
\end{eqnarray}
where $\zh=\xh^1+i\xh^2,~\zbh=\xh^1-i\xh^2$, then
we get the Heisenberg's commutation relation:
\begin{eqnarray}
{[\ah,\ah^\dagger]}=1.
\end{eqnarray}
Hence the spatial coordinates can be considered 
as the operators acting on
Fock space $\cH$ which is spanned by the occupation number basis $\vert
n\ket:=\left\{(\ah^\dagger)^n/\sqrt{n!}\right\}\vert 0\ket,~\ah\vert 0\ket=0$:
\begin{eqnarray}
\label{fock}
\cH=\oplus_{n=0}^{\infty}\C\vert n\ket.
\end{eqnarray}

The fields on the space depend on the spatial coordinates 
and are also the operators
acting on the Fock space $\cH$. 
They are represented by the occupation number basis as
\begin{eqnarray}
\fh=\sum_{m,n=0}^{\infty}f^{(mn)}\vert m\ket\bra n\vert.
\end{eqnarray}
The matrix element $f^{(mn)}$ is infinite-size.
If the fields has rotational symmetry on the plane,
that is, the fields commute with the number operator $\nh:=\ah^\dagger 
\ah\sim (\xh^1)^2+(\xh^2)^2$,
they become diagonal:
\begin{eqnarray}
\fh=\sum_{n=0}^{\infty}f^{(n)}\vert n\ket\bra n\vert.
\end{eqnarray}

The derivative of an operator $\hat{\cO}$ can be defined by
\begin{eqnarray}
\del_\mu \hat{\cO} :=[\delh_\mu,\hat{\cO}],~~~~~~\mbox{where}~~~
\delh_\mu:=-i(\theta^{-1})_\mn \xh^\nu,
\end{eqnarray}
which satisfies Leibniz rule and
the desirable relation : $\del_\mu \xh^{\nu}
=\delta_{\mu}^{~\nu}$.
Moreover defining the following anti-Hermitian operator
\begin{eqnarray}
\Dh_\mu:=\delh_\mu+\Ah_\mu,
\end{eqnarray} 
where the $\Ah_\mu$ is a gauge field and anti-hermitian, then
the covariant derivative of an adjoint field $\Phih$ 
can be defined by $[\Dh_\mu,\Phih]$.

We note that using this anti-Hermitian operator $\Dh_\mu$,
 the field strength $\Fh_\mn$ is rewritten as
\begin{eqnarray}
\Fh_\mn=[\Dh_\mu,\Dh_\nu]-i(\theta^{-1})_\mn.
\end{eqnarray} 
Here the constant term $-i(\theta^{-1})_\mn$ appears so that
it should cancel out the term $[\hat{\del}_\mu,\hat{\del_\nu}]
(=i(\theta^{-1})_\mn)$ 
in $[\hat{D}_\mu,\hat{D}_\nu]$
and becomes an obstruct in applying
``solution generating technique''
to BPS equations.

From now on, we mainly use complex representations
such as $\Dh_z:=(1/2)(\Dh_1-i\Dh_2)=-(1/2\theta)\zbh+\Ah_z$.

\subsection{``Solution Generating Technique''}

``Solution generating technique'' is a transformation which leaves 
an equation as it is, that is, one of the auto-B\"acklund
transformations.
The transformation is almost a gauge transformation
and defined as follows:
\begin{eqnarray}
\label{hkl}
\Dh_z\rar \Uh^\dagger \Dh_z \Uh,
\end{eqnarray}
where $\Uh$ is an almost unitary operator and satisfies
\begin{eqnarray}
\label{isometry}
\Uh \Uh^\dagger=1.
\end{eqnarray}
We note that we don't put $\Uh^\dagger\Uh=1$.
If $\Uh$ is finite-size, $\Uh\Uh^\dagger=1$ 
implies $\Uh^\dagger \Uh=1$ and then $\Uh$ and 
the transformation (\ref{hkl}) become a unitary operator and
just a gauge transformation respectively.
Now, however, $\Uh$ is infinite-size 
and we only claim that $\Uh^\dagger\Uh$ is a projection 
because $(\Uh^\dagger\Uh)^2=\Uh^\dagger(\Uh \Uh^\dagger)\Uh=\Uh^\dagger\Uh$.
The operator $\Uh$ 
which satisfies $\Uh \Uh^\dagger=1$ and $\Uh^\dagger \Uh=$ (projection) 
is often called the partial isometry. 

The transformation (\ref{hkl}) generally 
leaves an equation of motion as it is \cite{HaKrLa}:
\begin{eqnarray}
\fr{\delta\cL}{\delta\cO}\rar\Uh^\dagger\fr{\delta\cL}{\delta\cO}\Uh,
\end{eqnarray}
where $\cL$ and $\cO$ are the Lagrangian and the field in the
Lagrangian.
Hence if one prepares a known solution of 
the equation of motion $\delta\cL/\delta\cO=0$, then we can get various
new solution of it by applying the transformation (\ref{hkl}) to
the known solution.

The typical example of the partial isometry $\Uh$ is a shift operator.
In $U(1)$ gauge theory, one of the shift operators 
acting on the Fock space (\ref{fock}) is 
\begin{eqnarray}
\label{shift1}
\Uh_k=\sum_{n=0}^{\infty}\vert n\ket\bra n+k\vert,
\end{eqnarray}
which satisfies
\begin{eqnarray}
\Uh_k \Uh_k^\dagger=1,~~~
\Uh_k^\dagger  \Uh_k=1-\Ph_k,
\end{eqnarray}
where $\Ph_k$ is a projection onto the $k$-dimensional subspace of
the Fock space $\cH$ and expressed as
\begin{eqnarray}
\label{proj}
\Ph_k:=\sum_{m=0}^{k-1}\vert m\ket\bra m\vert.
\end{eqnarray}
Here we note that in star product 
formalism, the behavior of the shift operator 
at large $\vert x\vert$ is order 1 which is denoted by $\cO(1)$. 
The new soliton solutions from vacuum solutions 
are called localized solitons.
The dimension of the projection $\Ph_k$ in fact
represents the charge of the localized solitons.
In general, the new solitons generated from known solitons 
by ``solution generating technique'' are
the composite of known solitons and localized solitons.

``Solution generating technique'' (\ref{hkl}) can be generalized
so as to include moduli parameters.
In $U(1)$ gauge theory,
the generalized transformation becomes as follows:
\begin{eqnarray}
\label{hkl2}
\Dh_z\rar \Uh_k^\dagger \Dh_z \Uh_k-\sum_{m=0}^{k-1}
\fr{\bar{\alpha}_z^{(m)}}{2\theta}\vert m\ket\bra m\vert,
\end{eqnarray}
where $\alpha_z^{(m)}$ is an complex number
and represents the position of the $m$-th localized soliton.

\subsection{Localized Instantons}

Localized instantons are obtained by applying 
``solution generating technique'' (\ref{hkl2}) to
the BPS equations of 4-dimensional noncommutative gauge theory.

First let's consider the 4-dimensional noncommutative space
with the coordinates $x^\mu,~\mu=1,2,3,4$ whose
noncommutativity is introduced as the canonical form:
\begin{eqnarray}
\label{can_nc_coord}
\theta^{\mn}=\left(
\ba{cccc}
0&\theta_1&0&0\\
-\theta_1&0&0&0\\
0&0&0&\theta_2\\
0&0&-\theta_2&0\\
\ea
\right).
\end{eqnarray}
The fields on the 4-dimensional noncommutative space whose
noncommutativity is (\ref{can_nc_coord})
are operators acting on Fock space $\cH=\cH_1\ot\cH_2$ 
where $\cH_1$ and $\cH_2$ are defined by the
same steps as the previous discussion 
corresponding to noncommutative $x_1$-$x_2$ plane
and 
noncommutative $x_3$-$x_4$ plane
respectively.
The element in the Fock space $\cH=\cH_1\ot\cH_2$ is denoted 
by $\vert n_1\ket\ot\vert n_2\ket$ or $\vert n_1,n_2\ket$.
We introduce the complex coordinates as $z_1=x_1+ix_2,~z_2=x_3+ix_4$.

Here
we make the noncommutative parameter $\theta^\mn$ anti-self-dual:
$\theta_1=-\theta_2=:\theta>0$,
so that ``solution generating technique'' could work well on
the BPS equation which is discussed later soon.
In this case, we can define annihilation operators 
as $\ah_1:=(1/\sqrt{2\theta})\zh_1,~\ah_2:=(1/\sqrt{2\theta})\zbh_2$
and creation operator $\ah_1^\dagger:=(1/\sqrt{2\theta})\zbh_1,~\ah_2^\dagger
:=(1/\sqrt{2\theta})\zh_2$ 
in Fock space $\cH=\oplus_{n_1,n_2=0}^{\infty}
\C\vert n_1\ket\otimes \vert n_2\ket$ such as 
\begin{eqnarray}
[\ah_1,\ah_1^\dagger]=1,~[\ah_2,\ah_2^\dagger]=1,~\mbox{otherwise}=0,
\end{eqnarray}
where $\vert n_1\ket$ and $\vert n_2\ket$ are the occupation number basis
generated form the vacuum state $\vert 0\ket$
by the action of $\ah_1^\dagger$ and $\ah_2^\dagger$ 
respectively.

4-dimensional noncommutative gauge theory is defined by 
the pure Yang-Mills action:
\begin{eqnarray}
\cL_{\scr\mbox{YM}}
=-\fr{1}{4g^2_{\scr\mbox{YM}}}
\int d^4x~\Tr F_\mn F^\mn,
\end{eqnarray}
where $\int d^4x$ denotes $\Tr_{\cH}$.

The anti-self-dual BPS equations are obtained as the condition
that the action density should take 
the minimum:
\begin{eqnarray}
\label{bps_instanton}
(\Fh_{z_1\zb_1}+\Fh_{z_2\zb_2}=)&&
-[\Dh_{z_1},\Dh_{z_1}^\dagger]-[\Dh_{z_2},\Dh_{z_2}^\dagger]
-\half\left(\fr{1}{\theta_1}+
\fr{1}{\theta_2}\right)=0,\no\\
(\Fh_{z_1z_2}=)&&[\Dh_{z_1},\Dh_{z_2}]=0,
\end{eqnarray}
The fields are denoted by the occupation number basis as
\begin{eqnarray}
\Ah_\mu(\xh)
&=&\sum_{m_1,m_2,n_1,n_2=0}^{\infty}c_\mu^{(m_1,m_2,n_1,n_2)}
\vert m_1,m_2\ket\bra n_1,n_2\vert\no\\
&=&\sum_{m_1,m_2,n_1,n_2=0}^{\infty}c_\mu^{(m_1,m_2,n_1,n_2)}
\vert m_1\ket\bra n_1\vert 
\ot \vert m_2\ket\bra n_2\vert,
\end{eqnarray}
where $c_\mu^{(m_1,m_2,n_1,n_2)}$ is a number.
We note that only when noncommutative parameter $\theta^{ij}$
is anti-self-dual, 
the constant term $\left(1/\theta_1+1/\theta_2\right)$ disappears 
and ``solution generating technique''
can leave the BPS equation (\ref{bps_instanton}) as it is.

Localized instanton solutions are generated by 
``solution generating technique'' from the vacuum solution
which trivially satisfies the BPS equation (\ref{bps_instanton})
and given by
\begin{eqnarray}
\label{loc_inst}
\Dh_{z_i}=\Uh_k^\dagger \delh_{z_i} \Uh_k-\sum_{m=0}^{k-1}
\fr{\bar{\alpha}_i^{(m)}}{2\theta_i}\vert 0,m\ket\bra 0,m\vert,
\end{eqnarray}
where the shift operators can be taken for example as \cite{Fu3}
\begin{eqnarray}
\label{shift2}
\Uh_k=\sum_{n_1=1,n_2=0}^{\infty}\vert n_1,n_2\ket\bra n_1,n_2\vert
+\sum_{n_2=0}^{\infty}\vert 0,n_2\ket\bra 0,n_2+k\vert,
\end{eqnarray}
which satisfies
\begin{eqnarray}
\Uh_k \Uh_k^\dagger=1,~~~
\Uh_k^\dagger  \Uh_k=1-\sum_{m=0}^{k-1}\vert 0\ket\bra 0
\vert\ot\vert m\ket\bra m\vert.
\end{eqnarray}
The field strength and the instanton number $\nu[\Ah]$ are calculated as
\begin{eqnarray}
\Fh_\mn&=&-i(\theta^{-1})_\mn\vert 0\ket\bra 0
\vert\ot\Ph_k,\\
\nu[\Ah]&:=&\fr{1}{16\pi^2}\int d^4x~\Fh_\mn\Fh^\mn
=-\dim_{\cH}\vert 0\ket\bra 0
\vert\ot\Ph_k
=-k
\end{eqnarray}
Therefore the existence of the non-trivial projection $\Ph_k$ is 
crucial in generating localized solitons 
and the dimension of the projection corresponds to
the instanton number.

The interpretation of the moduli parameter $\alpha_i^{(m)}$
is clear in commutative description.
The exact Seiberg-Witten map \cite{OkOo} 
of the solution (\ref{loc_inst}) is
obtained in \cite{HaOo} and the D0-brane density is
\begin{eqnarray}
\label{d0density}
J_{\scr\mbox{D0}}(x)=\fr{2}{\theta^2}
+\sum_{m=0}^{k-1}\delta(x_1-\lambda_1^{(m)})\delta(x_2-\lambda_2^{(m)})
\delta(x_3-\lambda_3^{(m)})\delta(x_4-\lambda_4^{(m)}),
\end{eqnarray}
where the real parameters $\lambda_\mu^{(m)}$ are the real or
the imaginary part of $\alpha_i^{(m)}$, 
that is, $\alpha_1^{(m)}=\lambda_1^{(m)}+i\lambda_2^{(m)},
~\alpha_2^{(m)}=\lambda_3^{(m)}+i\lambda_4^{(m)}$.
The first term and the second term of the right hand side in
(\ref{d0density}) show the uniform distribution of the D0-branes on
D4-brane and localized $k$-D0-brane charge respectively,
which represents just the $k$-localized instantons.
The moduli parameter $\alpha_i^{(m)}$ 
or $\lambda_\mu^{(m)}$ are clearly interpreted as
the positions of the localized instantons.

\subsection{BPS Fluxons}

BPS fluxons are obtained by applying 
``solution generating technique'' to 
the BPS equation of $(3+1)$-dimensional
noncommutative gauge theory
with the coordinates $(x^0,x^i),~i=1,2,3$ 
whose noncommutativity is $\theta^{12}=\theta>0$.

$(3+1)$-dimensional
noncommutative gauge theory
is defined by the Yang-Mills-Higgs action:
\begin{eqnarray}
I_{\scr\mbox{YMH}}
=-\fr{1}{4g^2_{\scr\mbox{YM}}}
\int d^4x~\Tr\left(\Fh_{\mn}\Fh^{\mn}
+2[\Dh_\mu,\Phih][\Dh_\mu,\Phih]\right),
\end{eqnarray}
where $\Phih$ is an adjoint Higgs field and $\int dx_1dx_2$ 
denotes $\Tr_{\cH}$.
The anti-self-dual BPS equations are obtained as subsection 2.2:
\begin{eqnarray}
\label{bps_monopole}
(\Bh_3=)&&[\Dh_z,\Dh_{z}^\dagger]+\fr{1}{\theta}=-[\Dh_3,\Phih],\no\\
(\Bh_z=)&&[\Dh_3,\Dh_{z}]=-[\Dh_z, \Phih],
\end{eqnarray}
where $\Bh_i$ are magnetic fields.
This equation is often called Bogomol'nyi equation \cite{Bo}.
The fields with rotational symmetry on $x_1$-$x_2$ plane 
are denoted by the occupation number basis as
\begin{eqnarray}
\Phih=\sum_{n=0}^{\infty}\Phi^{(n)}(x_3)\vert n\ket\bra n\vert,~~~
\Ah=\sum_{n=0}^{\infty}A^{(n)}(x_3)\vert n\ket\bra n\vert.
\end{eqnarray}

Because of the constant term in the left hand side of the first
equation of (\ref{bps_monopole}), ``solution generating technique''
(\ref{hkl2}) cannot work.
The modified ``solution generating technique'' which leaves
the BPS equation (\ref{bps_monopole}) as it is
is found in \cite{HaTe}, \cite{Hashi}:
\begin{eqnarray}
\label{ours}
\Phih&\rar& \Uh_k^\dagger\Phih \Uh_k-\fr{x_3}{\theta} \Ph_k
+\sum_{m=0}^{k-1}\lambda^{(m)}_{\Phi}
\vert m\ket\bra m \vert,\no\\
\Dh_3&\rar&\del_3+\Uh_k^\dagger\Ah_3\Uh_k
-i\sum_{m=0}^{k-1}\fr{\lambda^{(m)}_4}{\theta}
\vert m\ket\bra m \vert,\no\\
\Dh_z&\rar& \Uh^\dagger_k\Dh_z\Uh_k
-\sum_{m=0}^{k-1}\fr{\bar{\alpha}^{(m)}_z}{2\theta}
\vert m\ket\bra m \vert,
\end{eqnarray}
where $\Uh_k$ and $\Ph_k$ are the same as (\ref{shift1}) and 
(\ref{proj}) respectively.
The important modification is to add the linear term of $x_3$
to the transformation of the Higgs field $\Phih$.
The localized soliton solutions in this theory
are generated from the vacuum solution
by the transformation
(\ref{ours})
\begin{eqnarray}
\label{fluxon}
\Phih&=&-\sum_{m=0}^{k-1}
\left(\frac{x_3}{\theta}-\lambda_\Phi^{(m)}\right)
\vert m\ket\bra m \vert,\nonumber\\
\Dh_z&=&\Uh^\dagger_k\hat{\partial}_z \Uh_k
-\sum_{m=0}^{k-1}\fr{\alpha_z^{(m)}}{2\theta}\vert m\ket\bra m \vert,
~~~\Ah_3
=-i\sum_{m=0}^{k-1}\fr{\lambda_4^{(m)}}{\theta}
\vert m\ket\bra m \vert,\nonumber\\
\Bh_3&=&\frac{1}{\theta}\Ph_k,
~~~\Bh_1
=\Bh_2=0,
\end{eqnarray}
which is called the BPS fluxon \cite{Po}, \cite{GrNe2} 
because this is similar to a flux-tube rather than a monopole.

The D1-brane density in commutative side is obtained by Seiberg-Witten 
map in \cite{HaOo}:
\begin{eqnarray}
\label{d1density}
J_{\scr\mbox{D1}}(x)=\fr{1}{\theta}\delta(\Phi)
+\sum_{m=0}^{k-1}\delta(x_1-\lambda_1^{(m)})\delta(x_2-\lambda_2^{(m)})
\delta(\Phi+(x_3-\lambda_3^{(m)})/\theta).
\end{eqnarray}
Hence the parameter $\lambda_i^{(m)}$ shows the positions of BPS
fluxon and here we use the relation $\lambda_\Phi=\lambda_3/\theta$
(cf. figure \ref{fluxonfig}). 
We can take $\lambda_4^{(m)}=0$ because 
$\lambda_4^{(m)}$ doesn't appear in (\ref{d1density}) and has no
physical meaning.

\section{ADHM/Nahm Construction of Localized Solitons}

In this section, we first review ADHM construction of commutative instantons
and then apply it to localized instantons,
localized periodic instantons (=localized calorons),
localized doubly-periodic instantons and BPS fluxons. 
The procedure of the constructions
are the same as the commutative case
and gives rise to various exact BPS solitons straightforwardly.
The shift operators and moduli terms naturally appear
in ADHM construction of localized instantons,
and the linear term of $x_3$ in (\ref{ours})
is necessarily obtained in Nahm construction of BPS fluxons.
The localized calorons and the localized doubly-periodic instantons
are new solitons.

\subsection{A Review of ADHM Construction of Instantons and Calorons}

In this subsection, we discuss ADHM construction of commutative instantons.
First let's introduce the Euclidean 4-dimensional Pauli matrices:
\begin{eqnarray}
e_\mu:=(-i\sigma_i,1),~~~e_\mu^\dagger=(i\sigma_i,1),
\end{eqnarray}
which correspond to 
the basis of the quarternion 
as algebra : $e_ie_j=-\delta_{ij}+\epsilon_{ijk}e_k$
and also satisfy the following relations:
\begin{eqnarray}
e_\mu e_\nu^\dagger&=&\delta_\mn +i\eta_\mn^{i(-)}\ot\sigma_i,\no\\
e_\mu^\dagger e_\nu&=&\delta_\mn +i\eta_\mn^{i(+)}\ot\sigma_i.
\end{eqnarray}
Here $\eta_\mn^{i(\pm)}$ are called 't Hooft symbol 
and concretely represented as
\begin{eqnarray}
\eta^{i(\pm)}_\mn=\epsilon_{i\mn 4}\pm\delta_{i\mu}\delta_{\nu 4}
\mp\delta_{i\nu}\delta_{\mu 4}.
\end{eqnarray}
These symbols are anti-symmetric and (anti-)self-dual.
Next we define ``0-dimensional Dirac operator'' 
which is $(N+2k)\ti 2k$ matrix as
\begin{eqnarray}
\nah&:=&
\left(
\ba{c}
S\\
(x^\mu-T^\mu)\ot e^\mu
\ea
\right)
=\left(
\ba{cc}
J&I^\dagger\\
-i(z_2-B_2)&-i(\zb_1-B_1^\dagger)\\
-i(z_1-B_1)&i(\zb_2-B_2^\dagger)
\ea
\right),
\end{eqnarray}
where $S$ and $T_\mu$ are $N\times 2k$ and $k\times k$ matrices 
respectively and $T_\mu$ are Hermitian : $T_\mu^\dagger=T_\mu$. $I$ and $J$ 
are $k\times N$ and $N\times k$ matrices respectively
and $B_1:=T_1+iT_2,~B_2:=T_3+iT_4$.

The matrices satisfy the following relations which is equivalent to
that $\na^\dagger\na$ commute with Pauli matrices $\sigma_i$:
\begin{eqnarray}
\label{com_adhm}
&&{[B_1,B_1^\dagger]}+[B_2,B_2^\dagger]+II^\dagger-J^\dagger J
~(\equiv -[z_1,\zb_1]-[z_2,\zb_2])=0,\no\\
&&{[B_1,B_2]}+IJ=0,
\end{eqnarray}
which are called ADHM equations.
Moreover we have to
put another condition on the matrices
that $\na^\dagger \na$ is invertible,
which is in fact necessary in ADHM construction.

ADHM construction consists of the following three steps.
The first step is to solve the ADHM equations.
Next step is to solve the following ``0-dimensional Dirac equation''
in the background of the solution of ADHM eq. (\ref{com_adhm}):
\begin{eqnarray}
\label{com_Dirac_instanton}
\na^\dagger V=0,
\end{eqnarray}
where $V$ is $(N+2k)\ti N$ matrices and satisfies the normalization condition:
\begin{eqnarray}
\label{norm}
V^\dagger V=1,
\end{eqnarray}
and completeness condition\footnote{
This condition on noncommutative space is discussed 
in \cite{KiLeYa}, \cite{ChKhTr}}:
\begin{eqnarray}
\label{comp}
VV^\dagger=1-\na (\na^\dagger \na)^{-1}\na^\dagger.
\end{eqnarray}
which comes from the assumption that $\na^\dagger \na$ is invertible.
It is convenient to introduce the following decomposed matrices
of $V$ :
\begin{eqnarray}
\label{com_v_decomposed}
V=
\left(
\ba{c}
u\\
v
\ea
\right)
=
\left(
\ba{c}
u\\
v_1\\
v_2
\ea
\right),
\end{eqnarray}
where $u,v$ and $v_{1,2}$ are $N\times N,~2k\times N$ and $k\times N$ 
matrices respectively.
We note that $u$ and $v$ behave $\cO(1)$ and $\cO(r^{-1})$
at $r=\vert x\vert \rar \infty$ respectively \cite{CoGo}.
The final step is to construct the (anti-)self-dual gauge fields
using the solution $V$ of the ``0-dimensional Dirac equation''
(\ref{com_Dirac_instanton})
as follows:
\begin{eqnarray}
\label{com_gauge_instanton}
A_\mu&=&V^\dagger\del_{\mu}V
=u^\dagger\del_\mu u+v^\dagger\del_\mu v.
\end{eqnarray}
The field strength is calculated from the gauge fields:
\begin{eqnarray}
\label{proof}
F&=&dA+A\wedge A\no\\
&=&dV^\dagger \wedge dV+V^\dagger dV\wedge V^\dagger dV
=dV^\dagger \wedge dV-dV^\dagger V\wedge V^\dagger dV\no\\
&=&dV^\dagger(1-VV^\dagger)\wedge dV
=dV^\dagger\na (\na^\dagger\na)^{-1}\na^\dagger \wedge dV\no\\
&=&V^\dagger (d\na) (\na^\dagger\na)^{-1}\wedge (d\na^\dagger) V
=v^\dagger e_\mu dx^\mu (\na^\dagger\na)^{-1}
\wedge e^{\dagger}_\nu dx^\nu v\no\\
&=&iv^\dagger(\na^\dagger\na)^{-1}\eta^{(-)}_{\mu\nu}v dx^\mu \wedge dx^\nu.\\
F_{\mu\nu}&=&2iv^\dagger(\na^\dagger\na)^{-1}\eta^{(-)}_{\mu\nu}v.
\end{eqnarray}
Hence anti-self-dual gauge fields have been constructed.
In the last line of the equation (\ref{proof}),
we use the condition that $\na^\dagger \na$ should commute with
Pauli matrices.

\vspace{2mm}
\noindent
\unl{\it $G=SU(2)$ 't Hooft $k$ instantons}
\vspace{2mm}

Let's construct $G=SU(2)$ 't Hooft $k$-instanton solution
following the steps in ADHM construction.
The solution of ADHM equation (\ref{com_adhm}) 
is simply given for this instanton as follows: 
\begin{eqnarray}
S&=&
\left(
\ba{ccc}
\ba{cc}
\rho_1&0\\0&\rho_1
\ea
&
\cdots
&
\ba{cc}
\rho_k&0\\0&\rho_k
\ea
\ea
\right),\no\\
T_\mu&=&\diag_{m=0}^{k-1}(\lambda_\mu^{(m)}),
\end{eqnarray}
where the symbol ``$\diag$'' denotes diagonal sum 
and $\lambda_\mu^{(m)}$ and $\rho_m$ are real numbers. 
``0-dimensional Dirac equation'' (\ref{com_Dirac_instanton})
is also simply solved:
\begin{eqnarray}
V=\fr{1}{\sqrt{\cN}}\left(
\ba{c}
1\\
-((x^\mu-T^\mu)\ot e^\dagger_\mu)^{-1}S^\dagger
\ea
\right),
\end{eqnarray}
where the normalization factor $\cN$ 
is determined by normalization condition
(\ref{norm}) as
\begin{eqnarray}
\cN=1+\sum_{m=0}^{k-1}\fr{\rho_m^2}{\vert x-\lambda^{(m)}\vert^2},
\end{eqnarray}
and
\begin{eqnarray}
((x^\mu-T^\mu) \ot e^\dagger_\mu)^{-1}
=\diag_{m=0}^{k-1}
\left(\fr{(x_\mu-\lambda_\mu^{(m)})}{\vert x-\lambda^{(m)}\vert^2}
\ot e^\mu\right).
\end{eqnarray}$\uh$ and $\vh$ are 
actually $\cO(1)$ and $\cO(r^{-1})$ respectively.
The gauge fields are given by
\begin{eqnarray}
\label{thooft}
A_\mu&=&V^\dagger\del_\mu V
=-\fr{i}{\cN}\sum_{m=0}^{k-1}
\fr{\rho_m^2\eta_\mn^{(+)}(x_\nu-\lambda_\nu^{(m)})}
{\vert x-\lambda^{(m)}\vert^4}
=-\fr{i}{2}\eta_\mn^{(+)}\del^{\nu}\log \cN.
\end{eqnarray}
This solution is called 't Hooft-instanton solution
and singular at $x=\lambda^{(m)}$, which results from 
that a singular gauge is taken.

\vspace{2mm}
\noindent
\unl{\it $G=SU(2)$ 1 caloron}
\vspace{2mm}

The solution (\ref{thooft}) 
can be generalized to periodic-instanton solution.
We can take the instanton number $k=\infty$ and all the size of the 
instantons $\rho_m=\rho$ and put them periodically
along the $x_4$ axis where the period is $\beta$.
This soliton is called the caloron \cite{HaSh} 
and then $\cN$ becomes
\begin{eqnarray}
\cN=1+\sum_{m=-\infty}^{\infty}\fr{\rho^2}{\vert x-m\beta x_4\vert^2}
=1+\fr{\pi \rho^2}{\beta\vert \vec{x}\vert}
\fr{\sinh\left(\dis\fr{2\pi}{\beta}\vert \vec{x}\vert\right)}
{\cosh\left(\dis\fr{2\pi}{\beta}\vert \vec{x}\vert\right)
-\cos\left(\dis\fr{2\pi}{\beta}x_4\right)},
\end{eqnarray}
where $\vec{x}=(x_1,x_2,x_3)$.

The caloron solution 
coincides with PS-monopole solution \cite{PrSo} up to gauge transformation
with $\beta\rar 0$.
PS-monopole solution is given by
\begin{eqnarray}
\Phi&=&-\fr{x^i\sigma_i}{\vert\vec{x}\vert^2}
\left(\fr{a\vert\vec{x}\vert}{\tanh a\vert\vec{x}\vert}-1\right),\no\\
A_i&=&\fr{\epsilon_{ijk}\sigma^j x^k}{\vert\vec{x}\vert^2}
\left(\fr{a\vert\vec{x}\vert}{\sinh a\vert\vec{x}\vert}-1\right),
\end{eqnarray}
where the real constant $a$
represents the vacuum expectation value of the Higgs field, which 
appears in the gauge transformation. This is reinterpreted clearly 
from D-brane picture in \cite{LeYi}. (cf. figure \ref{caloron}.)
We will discuss the similar discussion about localized caloron
solution in section 4.

\subsection{ADHM Construction of Localized Instantons and Calorons}

Now let's generalize the above discussion to noncommutative case.
The difference to commutative case is that the coordinates are
operators which act on the Fock space. 
ADHM equation is deformed by the noncommutativity of the spatial
coordinates as follows:
\begin{eqnarray}
\label{adhm}
&&{[B_1,B_1^\dagger]}+[B_2,B_2^\dagger]+II^\dagger-J^\dagger J
=-2(\theta_1+\theta_2),\no\\
&&{[B_1,B_2]}+IJ=0.
\end{eqnarray}
We note that the constant term
in the right hand side of the first
equation disappears
only when the noncommutative parameter is
anti-self-dual, that is, $\theta_1+\theta_2=0$,
which is necessary for the existence of the localized instantons.

The steps to give rise to instantons are the same as the commutative case.

\vspace{2mm}
\noindent
\unl{\it localized $U(1)~k$ instantons}
\vspace{2mm}

Now let's find localized $U(1)$ instanton solutions using ADHM
construction, which is
considered as the noncommutative version of the 't Hooft instanton
solution in the $\rho^{(m)}\rar 0$ limit. 

ADHM equations (\ref{adhm}) are simply solved and
the solution of them for localized instantons is
\begin{eqnarray}
\label{sol_adhm}
I=J=0,~~~B_1=\diag_{m=0}^{k-1}(\alpha_1^{(m)}),~~~
B_2=\diag_{m=0}^{k-1}(\alpha_2^{(m)}),
\end{eqnarray}
where $\alpha_i^{(m)}$
should show the position of the $m$-th instanton
because $B_{1,2}$ is the scalar field on D0-branes. $I$ and $J$ 
contain the information of the
size of instantons and hence $I=J=0$ characterize the
localized instantons because localized instantons 
have no moduli parameter of the size and singular
on commutative side as (\ref{d0density}).

Next we solve ``0-dimensional Dirac equation'' 
in the background of the solutions (\ref{sol_adhm})
of the ADHM equation.
This is also simple.
Observing the right hand side of the complete condition (\ref{comp}),
we get $\vh_1^{(m)}
=\vert \alpha^{(m)}_1,\alpha^{(m)}_2\ket
\bra p_1^{(m)},p_2^{(m)}\vert$ and $\vh_2=0$,
where $\vert p_1^{(m)},p_2^{(m)}\ket$ is the 
normalized orthogonal state in $\cH_1\ot \cH_2$:
\begin{eqnarray}
\bra p_1^{(m)},p_2^{(m)}\vert p_1^{(n)},p_2^{(n)}\ket=\delta_{mn},
\end{eqnarray}
and $\vert \alpha^{(m)}_1,\alpha^{(m)}_2\ket$ is the 
normalized coherent state and satisfies
\begin{eqnarray}
\label{coherent}
&&\zh_1\vert \alpha^{(m)}_1,\alpha^{(m)}_2\ket
=\alpha^{(m)}_1\vert \alpha^{(m)}_1,\alpha^{(m)}_2\ket,\no\\
&&\zbh_2\vert \alpha^{(m)}_2,\alpha^{(m)}_2\ket
=\bar{\alpha}^{(m)}_2\vert \alpha^{(m)}_1,\alpha^{(m)}_2\ket,\no\\
&&\bra\alpha^{(m)}_1,\alpha^{(m)}_2\vert \alpha^{(m)}_1,\alpha^{(m)}_2\ket=1.
\end{eqnarray}
The eigen values $\alpha^{(m)}_1$ and $\alpha_2^{(m)}$
of $\zh_1$ and $\zbh_2$ are decided to be just the same as the $m$-th
diagonal components of the solution $B_1,B_2$ in (\ref{sol_adhm}).
Though $\uh$ is undetermined, $\Vh$ already satisfies $\na^\dagger
\Vh=0$, which comes from that 
in the case that the self-duality of gauge fields and
noncommutative parameter are the same, the coordinates in each column
of $\na^\dagger$
play the same role in the sense that they are annihilation operators or 
creation operators.

The last condition is the
normalization condition (\ref{norm})
and determines $\uh=\Uh_k$ where
$\Uh_k\Uh_k^\dagger=1,~
\Uh_k^\dagger \Uh_k=1-\Ph_k=1
-\sum_{m=0}^{k-1}\vert p_1^{(m)},p_2^{(m)}\ket\bra
p_1^{(m)},p_2^{(m)}\vert$.
This is just the shift operator and naturally appears in this way.
The shift operator and $\uh$ have the same behavior 
at $\vert x\vert\rar\infty$ and this is consistent.

Gathering the results, 
the solution of (\ref{com_Dirac_instanton}) is
\begin{eqnarray}
\label{v_instanton}
\Vh=
\left(\ba{c}
\uh\\
\vh_1^{(m)}\\
\vh_2^{(m)}
\ea\right)
=
\left(
\ba{c}
\Uh_k\\
\vert \alpha^{(m)}_1,\alpha^{(m)}_2\ket\bra p_1^{(m)},p_2^{(m)}\vert\\
0
\ea\right),
\end{eqnarray}
where $\vh_i^{(m)}$ is the $m$-th low of $\vh_i$.
This is the general form of the solution 
of ``0-dimensional Dirac equation'' and
gives rise to localized instanton solution:
\begin{eqnarray}
\Ah_{z_i}&=&\Vh^\dagger [\delh_{z_i},\Vh]
=\uh^\dagger \delh_{z_i} \uh+\vh^\dagger \delh_{z_i} \vh
-\delh_{z_i}\no\\
&=&\Uh_k^\dagger \delh_{z_i} \Uh_k- \vert p_1^{(m)},p_2^{(m)}\ket
\bra \alpha^{(m)}_1,
\alpha^{(m)}_2\vert\fr{\zbh_i}{2\theta^i} \vert 
\alpha^{(m)}_1,\alpha^{(m)}_2\ket\bra p_1^{(m)},p_2^{(m)}
\vert-\delh_{z_i}\no\\
&=&\Uh_k^\dagger \delh_{z_i} \Uh_k-\delh_{z_i}-\sum_{m=0}^{k-1}
\fr{\bar{\alpha}_{z_i}^{(m)}}{2\theta^i}
\vert p_1^{(m)},p_2^{(m)}\ket\bra p_1^{(m)},p_2^{(m)}\vert.
\end{eqnarray}

If $\vert p_1^{(m)},p_2^{(m)}\ket=\vert 0,m\ket$ and  $\Uh_k$ is 
the same as (\ref{shift2}), then the gauge fields
is the same as (\ref{loc_inst}).

The solution $\Vh$ of ``0-dimensional Dirac equation'' 
also contains
all informations of the instantons.
The instanton number $k$
is represented
by the dimension of the projected states$\vert p_1^{(m)},p_2^{(m)}\ket$
which appears in the relations of the shift operator $\uh=\Uh_k$ 
or the bra part of $\vh_1^{(m)}$
The information of the position of $k$ localized solitons 
is shown in the coherent state $\vert\alpha_i^{(m)}\ket$ in the ket 
part of $\vh_1^{(m)}$. 

\vspace{2mm}
\noindent
\unl{\it localized $U(1)$ 1 caloron}
\vspace{2mm}

Now let's construct a localized caloron solution as commutative
caloron solution in subsection 3.1, that is, we take the instanton number 
$k\rar \infty$ and put
infinite number of localized instantons in the $x_4$ direction 
at regular intervals. We have to find an appropriate shift operator
so that it gives rise to an infinite-dimensional projection operator
and put the moduli parameter $\lambda_4$ periodic.
 
The solution is found as:
\begin{eqnarray}
\label{loc_caloron}
\Ah_{z_1}&=&\Uh_{k\ti\infty}^\dagger \delh_{z_1} \Uh_{k\ti\infty}-\delh_{z_1}
-\sum_{m=0}^{k-1}
\fr{\bar{\alpha}_1^{(m)}}{2\theta}\vert m\ket\bra m\vert\ot 1_{\cH_2},
\no\\
\Ah_{z_2}&=&\Uh_{k\ti\infty}^\dagger \delh_{z_2} \Uh_{k\ti\infty}-\delh_{z_2}
+\sum_{m=0}^{k-1}\sum_{n=-\infty}^{\infty}
\fr{\bar{\alpha}_2^{(m)}-in\beta}
{2\theta}\vert m\ket\bra m\vert\ot \vert n\ket\bra n\vert,
\end{eqnarray}
where the shift operator is defined as
\begin{eqnarray}
\Uh_{k\times \infty}
=\sum_{n_1=0}^{\infty}\vert n_1\ket\bra n_1+k\vert\ot1_{\cH_2}.
\end{eqnarray}
The field strength is calculated as 
\begin{eqnarray}
\label{f_cal}
\Fh_{12}=-\Fh_{34}=i\fr{1}{\theta}\Ph_k\ot 1_{\cH_2},
\end{eqnarray}
which is trivially periodic in the $x_4$ direction. 
It seems to be strange 
that  this contains no information of the period $\beta$.
Hence one may wonder if this solution is the charge-one caloron
solution on $\R^3\ti S^1$ whose perimeter is $\beta$.
Moreover one may doubt if
this suggests that this soliton represents D2-brane 
not infinite number of D0-branes.

The apparent paradox is solved by mapping this solution to commutative side
by exact Seiberg-Witten map.
The commutative description of D0-brane density
is as follows 
\begin{eqnarray}
\label{d0density2}
J_{\scr\mbox{D0}}(x)=\fr{2}{\theta^2}
+\sum_{m=0}^{k-1}\sum_{n=-\infty}^{\infty}
\delta(x_1-\lambda_1^{(m)})\delta(x_2-\lambda_2^{(m)})
\delta(x_3-\lambda_3^{(m)})
\delta(x_4-\lambda_4^{(m)}-n\beta).
\end{eqnarray}
The information of the period has appeared
and the solution (\ref{loc_caloron}) is 
shown to be an appropriate charge-one caloron solution
with the period $\beta$.
The above paradox is due to the fact that in noncommutative gauge theories,
there is no local observable and the period becomes obscure\footnote{
Without Seiberg-Witten map, we can discuss the physical meaning of
the moduli parameter $\lambda_\mu$ on noncommutative side, 
for example, see \cite{AgGoMiSt}, \cite{BaLePa}, \cite{GrNe3}.}.
And as is pointed out in \cite{HaOo},
the D2-brane density is exactly zero. 
Hence the paradox has been solved clearly.

This soliton can be interpreted as a localized instanton
on noncommutative $\R^3\ti S^1$.

\vspace{2mm}
\noindent
\unl{\it localized $U(1)$ 1 doubly-periodic instantons}
\vspace{2mm}

In similar way, we can construct doubly-periodic (in the $x_3$ and
$x_4$ directions) instanton solution:
\begin{eqnarray}
\label{double}
\Ah_{z_1}&=&\Uh_{k\ti\infty}^\dagger \delh_{z_1} \Uh_{k\ti\infty}-\delh_{z_1}
-\sum_{m=0}^{k-1}
\fr{\bar{\alpha}_1^{(m)}}{2\theta}\vert m\ket\bra m\vert\ot 1_{\cH_2},
\no\\
\Ah_{z_2}&=&\Uh_{k\ti\infty}^\dagger \delh_{z_2} \Uh_{k\ti\infty}-\delh_{z_2}
\no\\
&&+\sum_{m=0}^{k-1}\sum_{n_1,n_2=-\infty}^{\infty}
\fr{\bar{\alpha}_2^{(m)}+\beta_1n_1-i\beta_2n_2}
{2\theta}\vert m\ket\bra m\vert\ot \vert 
\widetilde{\alpha}_{n_1n_2}^{(l_1,l_2)}
\ket\bra \widetilde{\alpha}_{n_1n_2}^{(l_1,l_2)}\vert,
\end{eqnarray}
where the system
$\left\{\vert \widetilde{\alpha}^{(l_1,l_2)}_{n_1,n_2}
\ket\right\}_{n_1,n_2\in\scr\Z}$
is von Neumann lattice \cite{vN} and an orthonormal and
complete set \cite{Pe}, \cite{BaBuGiKl}\footnote{
To make this system complete, the sum over the labels $(n_1,n_2)$ 
of von Neumann lattice is taken removing some one pair.
We apply this summation rule 
to the doubly-periodic instanton solution (\ref{double}).}.
Von Neumann lattice is the complete subsystem of the set of
the coherent states which is over-complete,
and generated by $e^{l_1\delh_3}$ and $e^{l_2\delh_4}$, 
where the periods of the lattice $l_1,l_2 \in \R$ 
satisfies $l_1 l_2=2\pi\theta$.
(See also \cite{BaGrZa}, \cite{GoHeSp}.)
This complete system has two kind of labels and suitable to
doubly-periodic instanton.
Of course, another complete system can be available if one label the
system appropriately. 

The field strength in the noncommutative side
is the same as (\ref{f_cal}) and 
the commutative description of D0-brane density becomes
\begin{eqnarray}
\label{d0density3}
J_{\scr\mbox{D0}}(x)&=&\fr{2}{\theta^2}
+\sum_{m=0}^{k-1}\sum_{n_1,n_2=-\infty}^{\infty}
\delta(x_1-\lambda_1^{(m)})\delta(x_2-\lambda_2^{(m)})\ti\no\\
&&~~~~~~~~~~~~~~~~~~~~~~~\delta(x_3-\lambda_3^{(m)}-n_1\beta_1)
\delta(x_4-\lambda_4^{(m)}-n_2\beta_2),
\end{eqnarray}
which
guarantees that this is an appropriate charge-one doubly-periodic
instanton solution
with the period $\beta_1,\beta_2$.

This soliton can be interpreted as a localized instanton on 
noncommutative $\R^2\ti T^2$.
The exact known solitons on noncommutative torus are very
refined or abstract as is found in
\cite{GoHeSp}, \cite{Boca}, \cite{KrSc}, \cite{KaMaTa}.
It is therefore notable that our simple solution (\ref{double}) is 
indeed doubly-periodic. The point is that
we treat noncommutative $\R^4$ not noncommutative torus
and apply ``solution generating technique'' to $\cH_1$ side
only. 

\vspace{2mm}
\noindent
\unl{\it localized $U(N)~k$ instantons}
\vspace{2mm}

There is an obvious generalization of the construction of $U(N)$
localized instanton as follows.
In the solution of ADHM equations, $I,J$ can be still zero 
and $B_{1,2}$ are the same as that of $N=1$ case.
The solution of ``0-dimensional Dirac equation'' (\ref{com_Dirac_instanton}) 
is given by
\begin{eqnarray}
\Vh=
\left(\ba{c}
\uh\\
\vh_1^{(m,a)}\\
\vh_2^{(m,a)}
\ea\right)
=
\left(
\ba{c}
\Uh_k\\
\vert \alpha^{(m_a)}_1,\alpha^{(m_a)}_2\ket\bra p_1^{(m_a)},p_2^{(m_a)}\vert 
\\
0
\ea\right),
\end{eqnarray}
where $m_a$ runs over some elements in
$\left\{0,1,\cdots,k-1\right\}$ whose number is $k_a$
and all $m_a$ are different. (Hence $\sum_{a=1}^N k_a=k$.) 
The $N\times N$ matrix $\Uh_k$ is a partial isometry and satisfies
\begin{eqnarray}
\Uh_k\Uh_k^\dagger=1,~~~
\Uh_k^\dagger \Uh_k=1-\Ph_k,
\end{eqnarray}
where the projection $\Ph_k$ is the following diagonal sum:
\begin{eqnarray}
\Ph_k:=\diag_{a=1}^{N}\left(\diag_{m_a}\vert p_1^{(m_a)},p_2^{(m_a)}\ket
\bra p_1^{(m_a)},p_2^{(m_a)}\vert\right).
\end{eqnarray}
$\vert\alpha^{(m_a)}_i\ket$ is the normalized coherent state (\ref{coherent}).

Next 
in the case of $\vert p_1^{(m_a)},p_2^{(m_a)}\ket=\vert 0,m_a\ket$, 
then the shift operator
is, for example, chosen as the following diagonal sum:
\begin{eqnarray}
\Uh_k=
\diag_{a=1}^{N}\left(
\sum_{n_1=1,n_2=0}^{\infty}\vert n_1,n_2\ket\bra n_1,n_2\vert
+\sum_{n_2=0}^{\infty}\vert 0,n_2\ket\bra 0,n_2+k_a\vert\right).
\end{eqnarray}
$\vert \alpha^{(m_a)}_1,\alpha^{(m_a)}_2\ket$ is the normalized coherent state
and defined similarly as (\ref{coherent}).
We can construct another non-trivial example of a shift operator
in $U(N)$ gauge theories by using noncommutative 
ABS construction \cite{AtBoSh}.
The localized instanton solution 
in \cite{Fu4} is one of these generalized solutions
for $N=2$.

We can construct $U(N)$ localized calorons and $U(N)$ localized
doubly-periodic instantons in the same way.

\subsection{Nahm Construction of BPS Fluxons}

In this subsection, 
we discuss the Nahm construction of $k$-BPS fluxon solutions.
The procedure is all the same as localized instantons.

In order to construct fluxon solution, we have to introduce
``1-dimensional Dirac operator'':
\begin{eqnarray}
\label{weyl_op_caloron}
\nah:=
\left(
\ba{cc}
J&I^\dagger\\
\dis i\fr{d}{d\xi}-i(x_3-T_3)&-i(\zbh_1-T_z^\dagger)\\
-i(\zh_1-T_z)&\dis i\fr{d}{d\xi}+i(x_3-T_3)
\ea
\right),
\end{eqnarray}
where $I,J$ and $T_\mu(\xi)$ are $k\times N,N\times k$ 
and $k\times k$ matrices
respectively and $T_\mu^\dagger=T_\mu,~T_z:=T_1+iT_2$.
We have taken the gauge $T_4=0$.

Now we introduce a formal product and an inner product of $N+2k$ 
vectors $\vec{V}(\xi)$ and $\vec{V}^{\pr}(\xi)$ as follows respectively
\begin{eqnarray}
\vec{V}\cdot\vec{V}^{\pr}&:=&\sum_{a=1}^{N}u_a^\dagger u_a^{\pr}
\delta(\xi-\xi_a)
+\vec{v}^\dagger\vec{v}^{\pr},\\
\bra \vec{V},\vec{V}^{\pr}\ket&:=&\int_{a_-}^{a_+}d\xi~
\vec{V}\cdot\vec{V}^{\pr}=\sum_{a=1}^{N}u_a^\dagger u_a^{\pr}
+\int_{a_-}^{a_+}d\xi~\vec{v}^\dagger\vec{v}^{\pr},
\end{eqnarray}
where $\vec{u}$ and $\vec{v}$ are the $N$ vector 
in the upper side of $\vec{V}$ and the $2k$ vector 
in the lower side of $\vec{V}$ respectively and 
$u_a$ is the $a$-th low of $\vec{u}$.
The components of $\vec{V}$ may contain differential operators.
The interval of integration in the inner product
depends on the kind of the monopoles
and is determined by the region
where the D1-brane spans in the transverse direction 
against the D3-branes (cf. figure \ref{caloron}).

The elements in the ``1-dimensional Dirac operator'' (\ref{weyl_op_caloron})
satisfy the following relation which is equivalent
to that $\nah\cdot\nah$ commutes with Pauli matrices $\sigma_i$:
\begin{eqnarray}
\label{nahm_monopole}
&&[T_z,T_z^\dagger]+[\fr{d}{d\xi}+T_3,-\fr{d}{d\xi}+T_3]
+\sum_{a=1}^{N}(I_aI_a^\dagger-J_a^\dagger J_a)\delta(\xi-\xi_a)
=-2\theta,\no\\
&&[T_z,\fr{d}{d\xi}+T_3]+\sum_{a=1}^{N}I_aJ_a\delta(\xi-\xi_a)=0. 
\end{eqnarray}
This is known as Nahm equation \cite{Nahm}\footnote{
Usually Nahm equation is written 
in the following real representation: $dT_i/d\xi+i\epsilon_{ijk}T_jT_k
+\sum_{a=1}^{N}S_a^\dagger S_a 
\delta(\xi-\xi_a)=-\theta\delta_{i3}$.}.
As in the case of instantons, the constant term appears 
in the right hand side of the first
equation because of the noncommutative parameters of the spatial coordinates.

If we define $\tilde{T}_i:=T_i+\theta\delta_{i3}\xi$,
the equation (\ref{nahm_monopole}) becomes
\begin{eqnarray}
\label{nahm_monopole2}
&&[\tilde{T}_z,\tilde{T}_z^\dagger]
+[\fr{d}{d\xi}+\tilde{T}_3,-\fr{d}{d\xi}+\tilde{T}_3]
+\sum_{a=1}^{N}(I_aI_a^\dagger-J_a^\dagger J_a)\delta(\xi-\xi_a)=0,\no\\
&&[\tilde{T}_z,\fr{d}{d\xi}+\tilde{T}_3]
+\sum_{a=1}^{N}I_aJ_a\delta(\xi-\xi_a)=0.
\end{eqnarray} 
This is the same as that on commutative space. 

Nahm construction also have three steps as ADHM construction,
that is, the first step is to solve the Nahm equation
(\ref{nahm_monopole}) and the next step is to solve 
the following ``1-dimensional Dirac equation''
in the background of the solution of Nahm equation
with the normalization condition:
\begin{eqnarray}
\label{weyl_monopole}
\nah\cdot\Vh
&=&
\sum_{a=1}^{N}
\left(
\ba{c}
J_a^\dagger\\I_a
\ea
\right)\uh_a\delta(\xi-\xi_a)\no\\
&&+
\left(
\ba{cc}
\dis i\fr{d}{d\xi}+i(x_3-T_3)&i(\zbh_1-T_{z}^\dagger)\\
i(\zh_1-T_z)&\dis i\fr{d}{d\xi}-i(x_3-T_3)
\ea
\right)
\left(\ba{c}\vh_1\\\vh_2\ea\right)
=0,\\
\label{norm_monopole}
\bra \Vh,\Vh\ket&=&1.
\end{eqnarray}
The third step is to construct  
the anti-self-dual configuration of Higgs field $\Phih$ and 
gauge fields $\Ah_i$ as follows:
\begin{eqnarray}
\label{gauge_monopole}
\Phih=\bra \Vh,\xi\Vh \ket,~~~
\Ah_i=\bra \Vh,\del_i\Vh \ket.
\end{eqnarray}
%
In the solution of the Higgs field, $\xi$ appears in place of
derivative, which suggests that the Higgs field would be
the Fourier-transformed
field of the gauge field $\Ah_4$.


Now let's construct BPS $k$-fluxon solution.
We put $G=U(1)$ and the coordinate of the jumping point $\xi_1=0$
for simplicity.
The situation is shown in figure \ref{fluxonfig}.

Nahm equation (\ref{nahm_monopole}) or (\ref{nahm_monopole2}) 
are simply solved similarly to ADHM equation:
\begin{eqnarray}
\label{sol_nahm_mono}
I=J=0,~~~
T_i(\xi)=\diag_{m=0}^{k-1}(\lambda_i^{(m)}-\theta\delta_{i3}\xi).
\end{eqnarray}
In fact $I$ and $J$ contain the information of the 
interval at the 
jumping points $\xi=\xi_a$ and $I=J=0$ shows that the interval $D=0$
(see figure \ref{fluxonfig}), which corresponds to BPS fluxons.

Next we have to solve the ``1-dimensional Dirac equation''
(\ref{weyl_monopole}).
We note that the interval of integration in the inner product
$\bra~,~\ket$ is infinite: $(-\infty,\infty)$ because the fluxon 
is described as the infinite D1-brane piercing D3-branes.

In the similar way of the instantons, 
the solution of Dirac equation (\ref{weyl_monopole}) in the background 
of (\ref{sol_nahm_mono}) can be found as follows:
\begin{eqnarray}
\label{v_monopole}
\Vh=
\left(
\ba{c}
\uh\\\vh_1^{(m)}\\\vh_2^{(m)}
\ea
\right)
=
\left(
\ba{c}
\Uh_k\\
f^{(m)}(\xi,x_3)
\vert \alpha_z^{(m)}\ket\bra m\vert\\
0
\ea
\right),
\end{eqnarray}
where $\vert \alpha_z^{(m)}\ket$ is the same as $\vert
\alpha_1^{(m)}\ket$ in subsection 2.2 and 
the partial isometry $\Uh_k$ is the same as (\ref{shift1}).

The function $f^{(m)}(\xi,x_3)$ is determined by  
the normalization condition (\ref{norm_monopole}) 
of $\Vh$ and ``1-dimensional Dirac equation''
(\ref{weyl_monopole}) as 
\begin{eqnarray}
\label{f}
f^{(m)}(\xi,x_3)=\left(\fr{\pi}{\theta}\right)^{\qua}
\exp\left[-\fr{\theta}{2}\left(\xi+\fr{x_3-\lambda_3^{(m)}}{\theta}\right)^2
\right].
\end{eqnarray}

Substituting (\ref{v_monopole}) and (\ref{f})
into (\ref{gauge_monopole}), 
we have the anti-self-dual configuration:
\begin{eqnarray}
\label{sol_monopole}
\Phih&=&
\xi_1\Uh_k^\dagger \Uh_k+\left(\fr{\theta}{\pi}\right)^{\half}
\sum_{m=0}^{k-1}\int_{-\infty}^{\infty}d\xi~
\left(\xi-\fr{x_3-\lambda_3^{(m)}}{\theta}
\right)e^{-\theta\xi^2}
\vert m\ket\bra m\vert\no\\
&=&-\sum_{m=0}^{k-1}\left(
\fr{x_3-\lambda_3^{(m)}}{\theta}\right)\vert m\ket\bra m\vert\no\\
\Ah_3&=&\bra\Vh,\del_3 \Vh\ket
=\int_{-\infty}^{\infty}d\xi~\vh^\dagger 
\left(-\fr{x_3-\lambda_3^{(m)}}{\theta}-\xi\right)\vh
=\sum_{m=0}^{k-1}\left(-\fr{x_3-\lambda_3^{(m)}}{\theta}-\Phi^{(m)}\right)
\vert m\ket\bra m\vert\no\\
&=&0,\no\\
\Ah_{z}&=&\bra\Vh,\del_{z}\Vh\ket
=\Uh_k^\dagger\hat{\del}_{z}\Uh_k-\hat{\del}_{z}
-\sum_{m=0}^{k-1}\fr{\bar{\alpha}_z^{(m)}}{2\theta}
\vert m\ket\bra m\vert.
\end{eqnarray}
This is just the BPS fluxon solution (\ref{fluxon}).
The linear term of $x_3$ in the Higgs field (\ref{fluxon})
is naturally derived and the meaningless parameter $\lambda_4^{(m)}$
of course never appears.

\section{Fourier Transformation of Localized Calorons}

In this section, we discuss the Fourier transformation
of the gauge fields of localized caloron
and show that the transformed configuration exactly coincides with
the BPS fluxon in the $\beta\rar 0$ limit.
This discussion is similar to that 
the commutative caloron exactly 
coincides with PS monopole in the $\beta\rar 0$ limit
up to gauge transformation as in the end of subsection 3.1,.

\begin{center}
\begin{figure}
\epsfxsize=110mm
\hspace{3cm} 
\epsffile{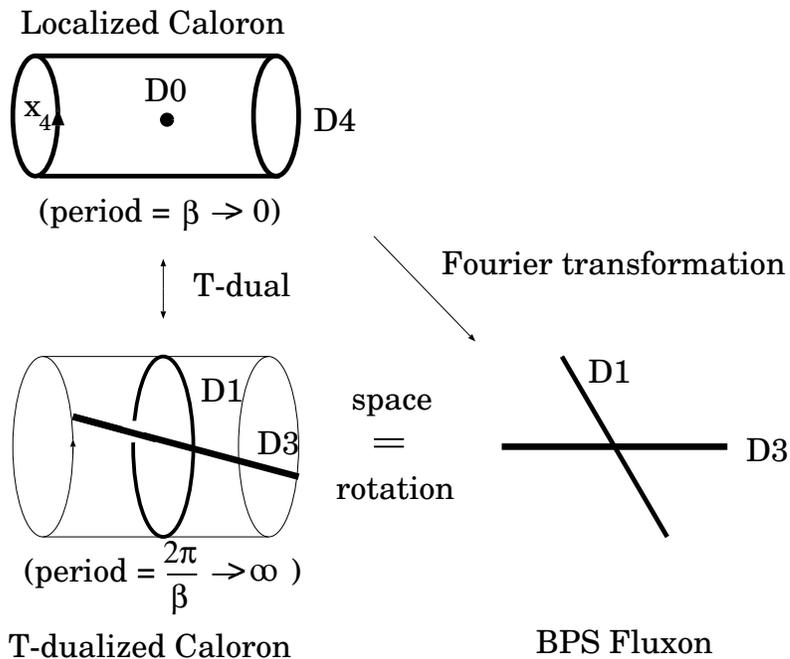}
\caption{Localized $U(1)$ 1 caloron 
and the relation to BPS fluxon}
\label{t-dual}
\end{figure}
\end{center}

The Fourier transformation can be defined as 
\begin{eqnarray}
\label{fourier}
\hat{1}_{\cH_2}&\rar& 1,~~~\xh_{3,4}\hat{1}_{\cH_2}\rar x_{3,4},\no\\
\Ah_\mu &\rar&\widetilde{\Ah_\mu^{[l]}}=\lim_{\beta\rar 0}\fr{1}{\beta}
\int_{-\fr{\beta}{2}}^{\fr{\beta}{2}}dx_4~e^{2\pi il\fr{x_4}{\beta}}\Ah_\mu.
\end{eqnarray}
In the  $\beta\rar 0$ limit, only $l=0$ mode survives and the
Fourier transformation (\ref{fourier}) becomes trivial.
Then we rewrite these zero modes $\widetilde{\Ah_i^{[0]}}$
and $i\widetilde{\Ah_4^{[0]}}$ as
$\Ah_i$ and $\Phih$ in $(3+1)$-dimensional
noncommutative gauge theory respectively.
Noting that in the localized caloron solution (\ref{loc_caloron}),
$\Uh_{k\ti\infty}^\dagger \delh_{z_2} \Uh_{k\ti\infty}-\delh_{z_2}=
\Ph_k\ot\hat{1}_{\cH_2}(\zbh_2/2\theta_2)$, where the $\Ph_k$
is the same as the projection in (\ref{proj}),
the transformed fields are easily calculated as follows:
\begin{eqnarray}
\Ah_{z_1}&=&\Uh_k^\dagger \delh_{z_1}\Uh_k-\delh_{z_1}-\sum_{m=0}^{k-1}
\fr{\bar{\alpha}_{z_1}^{(m)}}{2\theta_1}\vert m\ket\bra m\vert,\no\\
\Ah_3&=&i\sum_{m=0}^{k-1}\fr{\lambda_4^{(m)}}{\theta_2}
\vert m\ket\bra m\vert,\no\\
\Phih&=&\sum_{m=0}^{k-1}\left(\fr{x_3-\lambda_3^{(m)}}{\theta_2}\right)
\vert m\ket\bra m\vert.
\end{eqnarray}
The Fourier transformation (\ref{fourier}) 
also reproduces the anti-self-dual
BPS fluxon
rewriting $\theta_1$, $\theta_2$ and $z_1$
as $\theta$, $-\theta$ and $z$ respectively.
We note that the anti-self-dual condition of the noncommutative parameter
$\theta_1+\theta_2=0$ in the localized caloron would
correspond to the anti-self-dual condition of the BPS fluxon.
In the D-brane picture,
the Fourier transformation (\ref{fourier}) can be considered 
as the composite
of T-duality in the $x_4$ direction and the space rotation
in $x_3$-$\Phi$ plane \cite{HaHi}, \cite{Moriyama2},
\cite{HaHiMo}. (cf. figure \ref{t-dual})

\section{Conclusion and  Discussion}

In this paper, we have discussed ADHM/Nahm construction of localized
solitons in noncommutative gauge theories and discuss Fourier
transformation of localized calorons.
We have found the various localized solitons 
including new solitons: localized calorons and localized 
doubly-periodic instantons.
The shift operators and the moduli terms naturally appear
in ADHM construction.
BPS fluxons are also
obtained straightforwardly by the steps of
Nahm construction without modifications or tricks.
The Fourier-transformed localized calorons exactly coincide
with BPS fluxons which is consistent with the T-dual picture
of the corresponding D-brane system up to space rotation.

One of further studies is 
the Nahm construction of exact non-Abelian caloron
solutions in noncommutative gauge theory
and the study of T-duality of the gauge fields
or more fundamentally the Dirac zero mode $\Vh$.
T-duality is usually studied not for
the fields on D-brane but for the metric or $B$-field.
However T-duality of the gauge fields
described by operator formalism is very important because
the formalism is suitable to deal with algebraically
and the study might be a key point
of noncommutative ADHM or Nahm
duality and noncommutative Nahm transformation 
on non-commutative 4-torus \cite{AsNeSc}.
If we find some concrete representation of Nahm transformation,
we must be able to reveal many aspects on it.

Another direction is the completion of noncommutative ADHM or Nahm
duality. 
One-to-one correspondence between instanton/monopole solutions
and ADHM/Nahm data up to gauge equivalence is rather trivial
from the D-brane picture with background constant $B$-field.
Nevertheless the study is worthwhile because the detailed D-brane
interpretation of noncommutative ADHM/Nahm
duality might be useful for finding higher dimensional
ADHM/Nahm constructions which corresponds to D0-D6 system or
D0-D8 system
with appropriate background constant $B$-field 
\cite{Witten2}, \cite{Ohta_k}
\footnote{For some discussions including these systems 
with background constant $B$-field, see \cite{D6D8}.}.
In these system, the existence of the $B$-field is important
to make the systems BPS and hence noncommutative gauge
theoretical description of them which is equivalent to the D-brane 
system might give rise to some hints toward exact solution 
in higher dimensional gauge theories.

What plays crucial role to generate noncommutative solitons
is shift operators and projection operators. In this paper, 
we find appropriate operators in each situation and discuss where they
appear in ADHM/Nahm construction.
On noncommutative 4-torus, however, it is difficult to find
such operators in terms of concrete representation of some basis
in the Fock space and we seem to have to use Morita equivalence
as in \cite{KaMaTa}. The relation between the 
localized doubly-periodic instanton solution 
(\ref{double}) in our notation and the solution in 
\cite{GoHeSp}, \cite{Boca}, \cite{KrSc}, \cite{KaMaTa} is interesting.
 
\vskip7mm\noindent
{\bf Acknowledgments}

\noindent
It is a great pleasure to thank 
Y.~Matsuo
for careful reading of the present manuscript,
S.~Terashima
for clarifying comments and inspiring conversations,
and
K.~Hosomichi
for patient, fruitful discussions and continuous encouragement. 
I would like to thank
K.~Furuuchi, H.~Kajiura, K.~Hashimoto and T.~Takayanagi
for helpful comments.
I am also grateful to
E.~Corrigan,
T.~Eguchi, T.~Hirayama,
Mr.~Ishibashi, K.~Ishikawa,
T.~Kawano, Kimyeong~Lee, S.~Moriyama,
Y.~Sugawara,
and T.~Watari
for related discussions.
The author thanks 
Summer Institute 2001 at Yamanashi, Japan, and 
the YITP at Kyoto University, 
where he had an invited review talk
``Recent Developments in Non-Commutative Gauge Theory''
during the YITP workshop
YITP-W-01-04 on ``QFT2001,'' 
and their organizers
for giving him 
valuable opportunities of discussions related for this work.
This work was supported in part by
the Japan Securities Scholarship Foundation (\#12-3-0403).


\begin{thebibliography}{99}

\bibitem{CoDoSc}
A.~Connes, M.~R.~Douglas and A.~Schwarz,
JHEP {\bf 9802} (1998) 003
[hep-th/9711162].

\bibitem{DoHu}
M.~R.~Douglas and C.~Hull,
JHEP {\bf 9802} (1998) 008
[hep-th/9711165].

\bibitem{SeWi}
N.~Seiberg and E.~Witten,
JHEP {\bf 9909} (1999) 032
[hep-th/9908142].

\bibitem{Harvey}
J.~A.~Harvey,
``Komaba lectures on noncommutative solitons and D-branes,''
hep-th/0102076.

\bibitem{NeSc}
N.~Nekrasov and A.~Schwarz, 
Commun.\ Math.\ Phys.\ {\bf 198} (1998) 689
[hep-th/9802068].

\bibitem{Fu}
K.~Furuuchi,
Prog.\ Theor.\ Phys.\ {\bf 103} (2000) 1043 
[hep-th/9912047].

\bibitem{MaMiMoSt}
M.~Marino, R.~Minasian, G.~Moore and A.~Strominger,
JHEP {\bf 0001} (2000) 005
[hep-th/9911206].

\bibitem{Tera}
 S.~Terashima,
Phys. Lett. {\bf B477} (2000) 292 
[hep-th/9911245].

\bibitem{Moriyama}
S.~Moriyama,
JHEP {\bf 0008} (2000) 014
[hep-th/0006056].

\bibitem{GrNe}
D.~J.~Gross and N.~A.~Nekrasov, 
JHEP {\bf 0007} (2000) 034
[hep-th/0005204].

\bibitem{Po}
A.~P.~Polychronakos,
Phys.\ Lett.\ B {\bf 495} (2000) 407
[hep-th/0007043].

\bibitem{GrNe2}
D.~J.~Gross and N.~A.~Nekrasov,
JHEP {\bf 0010} (2000) 021 
[hep-th/0007204].

\bibitem{Nakajima}
H.~Nakajima,
``Resolutions of moduli spaces of ideal instantons on $\R^4$,''

{\it Topology, Geometry and Field Theory} 
(1994) 129 [ISBN/981-02-1817-6];

H.~Nakajima,
{\it Lectures on Hilbert Schemes of Points on Surfaces} 
(1999) [ISBN/0-8218-1956-9].

\bibitem{AgGoMiSt}
M.~Aganagic, R.~Gopakumar, S.~Minwalla and A.~Strominger,
JHEP {\bf 0104} (2001) 001
[hep-th/0009142].

\bibitem{Nek}
N.~A.~Nekrasov,
``Noncommutative instantons revisited,''
hep-th/0010017.

\bibitem{Fu4}
K.~Furuuchi,
JHEP {\bf 0103} (2001) 033
[hep-th/0010119].

\bibitem{AtHiDrMa}
M.~F.~Atiyah, N.~J.~Hitchin, V.~G.~Drinfeld and Y.~I.~Manin,
Phys.\ Lett.\ A {\bf 65} (1978) 185;
V.~G.~Drinfeld and Yu.~I.~Manin, 
Commun.\ Math.\ Phys.\ {\bf 63} (1978) 177.

\bibitem{Nahm}
W.~Nahm,
Phys.\ Lett.\ B {\bf 90} (1980) 413;

W.~Nahm,
``The construction of all self-dual multimonopoles by the ADHM method,''
{\it Monopoles in Quantum Field Theory}
(1982) 87 [ISBN/9971-950-29-4].

\bibitem{Witten} 
E.~Witten, 
 Nucl.\ Phys.\ B {\bf 460} (1996) 541
[hep-th/9511030].

\bibitem{Douglas} 
M.~R.~Douglas, 
``Branes within branes,''
hep-th/9512077;\\
M.~R.~Douglas, 
J. Geom. Phys. {\bf 28} (1998) 255
[hep-th/9604198].

\bibitem{Diaconescu} 
D.~Diaconescu, 
Nucl.\ Phys.\  B {\bf 503} (1997) 220 
[hep-th/9608163].

\bibitem{Myers}
R.~C.~Myers,
JHEP {\bf 9912} (1999) 022
[hep-th/9910053].

\bibitem{BaFiShSu}
T.~Banks, W.~Fischler, S.~H.~Shenker and L.~Susskind,
Phys.\ Rev.\  D {\bf 55} (1997) 5112
[hep-th/9610043].

\bibitem{IsKaKiTs}
N.~Ishibashi, H.~Kawai, Y.~Kitazawa and A.~Tsuchiya,
Nucl.\ Phys.\  B {\bf 498} (1997) 467
[hep-th/9612115].

\bibitem{AoIsIsKaKiTa}
H.~Aoki, N.~Ishibashi, S.~Iso, H.~Kawai, Y.~Kitazawa and T.~Tada,
Nucl.\ Phys.\ B {\bf 565} (2000) 176
[hep-th/9908141].

\bibitem{HaKrLa}
J.~A.~Harvey, P.~Kraus and F.~Larsen,
JHEP {\bf 0012} (2000) 024
[hep-th/0010060].

\bibitem{HaTe}
M.~Hamanaka and S.~Terashima,
JHEP {\bf 0103} (2001) 034
[hep-th/0010221].

\bibitem{Hashi}
K.~Hashimoto,
JHEP {\bf 0012} (2000) 023
[hep-th/0010251].

\bibitem{Bak}
D.~Bak,
Phys.\ Lett.\ B {\bf 495} (2000) 251
[hep-th/0008204].

\bibitem{BaLePa}
D.~Bak, K.~Lee and J.~H.~Park,
Phys.\ Rev.\ D {\bf 63} (2001) 125010
[hep-th/0011099].

\bibitem{HaSh}
B.~J.~Harrington and H.~K.~Shepard,
Phys.\ Rev.\ D {\bf 17} (1978) 2122;
Phys.\ Rev.\ D {\bf 18} (1978) 2990.

\bibitem{GrPiYa}
D.~J.~Gross, R.~D.~Pisarski and L.~G.~Yaffe,
Rev.\ Mod.\ Phys.\ {\bf 53} (1981) 43.

\bibitem{Ro}
P.~Rossi,
Nucl.\ Phys.\ B {\bf 149} (1979) 170.

\bibitem{LeYi}
K.~Lee and P.~Yi,
Phys.\ Rev.\ D {\bf 56} (1997) 3711
[hep-th/9702107].

\bibitem{Nahm2}
W.~Nahm,
``Self-dual monopoles and calorons,''
Lecture Notes in Physics {\bf 201} (1984) 189.

\bibitem{KrvB}
T.~C.~Kraan and P.~van Baal,
Phys.\ Lett.\ B {\bf 428} (1998) 268
[hep-th/9802049];
Nucl.\ Phys.\ B {\bf 533} (1998) 627
[hep-th/9805168].

\bibitem{LeLu}
K.~Lee and C.~Lu,
Phys.\ Rev.\ D {\bf 58} (1998) 025011
[hep-th/9802108].

\bibitem{PrSo} 
M.~K.~Prasad and C.~M.~Sommerfield, 
Phys.\ Rev.\ Lett.\ {\bf 35} (1975) 760.

\bibitem{HaHi}
K.~Hashimoto and T.~Hirayama,
Nucl.\ Phys.\ B {\bf 587} (2000) 207, 
[hep-th/0002090].

\bibitem{Moriyama2}
S.~Moriyama,
Phys.\ Lett.\  {\bf B485} (2000) 278
[hep-th/0003231].

\bibitem{HaHiMo}
K.~Hashimoto, T.~Hirayama and S.~Moriyama,
JHEP {\bf 0011} (2000) 014
[hep-th/0010026].

\bibitem{Fu3}
K.~Furuuchi,
``Topological charge of U(1) instantons on noncommutative $\R^4$,''
hep-th/0010006.

\bibitem{OkOo}
Y.~Okawa and H.~Ooguri,
Phys.\ Rev.\ D {\bf 64} (2001) 046009
[hep-th/0104036].

\bibitem{HaOo}
K.~Hashimoto and H.~Ooguri,
Phys.\ Rev.\ D {\bf 64} (2001) 106005
[hep-th/0105311].

\bibitem{Bo} 
E.~B.~Bogomol'nyi, 
Sov.\ J.\ Nucl.\ Phys.\ {\bf 24} (1976) 449.

\bibitem{KiLeYa}
K.~Y.~Kim, B.~H.~Lee and H.~S.~Yang,
``Comments on instantons on noncommutative $\R^4$,''
hep-th/0003093.

\bibitem{ChKhTr}
C.~S.~Chu, V.~V.~Khoze and G.~Travaglini,
``Notes on noncommutative instantons,''
hep-th/0108007.

\bibitem{CoGo} 
E.~Corrigan and P.~Goddard, 
Ann.\ Phys.\ {\bf 154} (1984) 253.

\bibitem{GrNe3}
D.~J.~Gross and N.~A.~Nekrasov,
JHEP {\bf 0103} (2001) 044
[hep-th/0010090].

\bibitem{vN}
J. von Neumann, 
{\it Mathematical Foundations of Quantum Mechanics}
(1996) [ISBN/0-691-02893-1].

\bibitem{Pe}
A. M. Perelomov,
Teor.\ Mat.\ Fiz.\ {\bf 6} (1971) 213.

\bibitem{BaBuGiKl}
V. Bargmann, P. Butera, L. Girardello and J. R. Klauder,
Rept.\ Math.\ Phys.\ {\bf 2} (1971) 221.

\bibitem{BaGrZa}
H. Bacry, A. Grossman and J. Zak,
Phys.\ Rev.\ B {\bf 12} (1975) 1118.

\bibitem{GoHeSp}
R.~Gopakumar, M.~Headrick and M.~Spradlin,
``On noncommutative multi-solitons,''
hep-th/0103256.

\bibitem{Boca}
F.~P.~Boca,
Commun.\ Math.\ Phys.\ {\bf 202} (1999) 325.

\bibitem{KrSc}
T.~Krajewski and M.~Schnabl,
JHEP {\bf 0108} (2001) 002
[hep-th/0104090].

\bibitem{KaMaTa}
H.~Kajiura, Y.~Matsuo and T.~Takayanagi,
JHEP {\bf 0106} (2001) 041
[hep-th/0104143].

\bibitem{AtBoSh}
M.~F.~Atiyah, R.~Bott and A.~Shapiro,
Topology {\bf 3} suppl. 1 (1964) 3. 

\bibitem{AsNeSc}
A.~Astashkevich, N.~Nekrasov and A.~Schwarz,
Commun.\ Math.\ Phys.\  {\bf 211} (2000) 167
[hep-th/9810147].

\bibitem{Witten2}
E.~Witten,
``BPS bound states of D0-D6 and D0-D8 systems in a B-field,''
hep-th/0012054.

\bibitem{Ohta_k}
K.~Ohta,
Phys.\ Rev.\ D {\bf 64} (2001) 046003
[hep-th/0101082].

\bibitem{D6D8}
B.~Chen, H.~Itoyama, T.~Matsuo and K.~Murakami,
Nucl.\ Phys.\ B {\bf 576} (2000) 177
[hep-th/9910263];
M.~Mihailescu, I.~Y.~Park and T.~A.~Tran,
Phys.\ Rev.\ D {\bf 64} (2001) 046006
[hep-th/0011079];
R.~Blumenhagen, V.~Braun and R.~Helling,
Phys.\ Lett.\ B {\bf 510} (2001) 311
[hep-th/0012157];
Y.~Imamura,
JHEP {\bf 0102} (2001) 035
[hep-th/0012254];
M.~Sato,
Int.\ J.\ Mod.\ Phys.\ A {\bf 16} (2001) 4069
[hep-th/0101226];
A.~Fujii, Y.~Imaizumi and N.~ Ohta,
Nucl.\ Phys.\ B {\bf 615} (2001) 61 [hep-th/0105079].

\end{thebibliography}
\end{document}